 \documentclass[final,5p,times,twocolumn]{elsarticle}
\pdfoutput=1

\usepackage{amssymb}
\usepackage{amsmath}
\usepackage{color}

\biboptions{numbers,square,comma,sort&compress}
\usepackage{url}
\journal{~}

\begin{document}
\begin{frontmatter}

%% Title, authors and addresses

%% use the tnoteref command within \title for footnotes;
%% use the tnotetext command for theassociated footnote;
%% use the fnref command within \author or \address for footnotes;
%% use the fntext command for theassociated footnote;
%% use the corref command within \author for corresponding author footnotes;
%% use the cortext command for theassociated footnote;
%% use the ead command for the email address,
%% and the form \ead[url] for the home page:
%% \title{Title\tnoteref{label1}}
%% \tnotetext[label1]{}
%% \author{Name\corref{cor1}\fnref{label2}}
%% \ead{email address}
%% \ead[url]{home page}
%% \fntext[label2]{}
%% \cortext[cor1]{}
%% \address{Address\fnref{label3}}
%% \fntext[label3]{}

\title{Constraints on neutrino decay lifetime using long-baseline charged and neutral current data}

%% use optional labels to link authors explicitly to addresses:
%% \author[label1,label2]{}
%% \address[label1]{}
%% \address[label2]{}

\author[Goias,Argonne]{R. A. Gomes} %\email{ragomes@if.ufg.br} 
\author[Goias]{A. L. G. Gomes}
\author[Unicamp,ICTP]{O. L. G. Peres}
\address[Goias]{Instituto de F\'isica, Universidade Federal de Goi\'as, 74001-970, Goi\^ania, GO, Brazil}
\address[Argonne]{Argonne National Laboratory, 60439, Argonne, Illinois, USA}
\address[Unicamp]{Instituto de F\'isica Gleb Wataghin, Universidade Estadual de Campinas, 13083-859, Campinas, SP, Brazil}
\address[ICTP]{Abdus Salam International Centre for Theoretical Physics, ICTP, I-34010, Trieste, Italy}

\begin{abstract}
%% Text of abstract

We investigate the status of a scenario involving oscillations and decay for charged and neutral current data from the MINOS and T2K experiments. We first present an analysis of charged current neutrino and anti-neutrino data from MINOS in the framework of oscillation with decay and obtain a best fit for non-zero decay parameter $\alpha_3$. The MINOS charged and neutral current data analysis results in the best fit for $|\Delta m_{32}^2| =  2.34\times 10^{-3}$~eV$^2$, $\sin^2 \theta_{23} = 0.60$ and zero decay parameter, which corresponds to the limit for standard oscillations. Our combined MINOS and T2K analysis reports a constraint at the 90\% confidence level for the neutrino decay lifetime $\tau_3/m_3 >  2.8 \times 10^{-12}$~s/eV. This is the best limit based only on accelerator produced neutrinos.
\end{abstract}

\begin{keyword}
%% keywords here, in the form: keyword \sep keyword
neutrino \sep neutrino decay \sep neutrino oscillation

%% PACS codes here, in the form: \PACS code \sep code
\PACS 14.60.Lm \sep 14.60.St \sep 14.60.Pq

%% MSC codes here, in the form: \MSC code \sep code
%% or \MSC[2008] code \sep code (2000 is the default)
\end{keyword}

\end{frontmatter}

%% \linenumbers

%% main text
\section{Introduction}
\label{intro}
Recent results from reactor experiments, Double Chooz~\cite{Abe:2012tg}, Daya Bay~\cite{An:2013zwz} and Reno~\cite{Ahn:2012nd},  complete the picture that three active neutrinos oscillate with two known non-zero mass differences 
($\Delta m_{21}^2\equiv m_2^2-m_1^2$ and $\Delta m_{31}^2\equiv m_3^2-m_1^2$), three mixing angles ($\theta_{12}$, $\theta_{23}$ and $\theta_{13}$)  and an unknown CP phase ($\delta$)~\cite{Pontecorvo:1957cp,Maki:1962mu}. For a detailed description of neutrino oscillations see Ref.~\cite{GonzalezGarcia:2002dz}.

In this picture the large statistics of atmospheric neutrinos of the Super-Kamiokande~\cite{Abe:2012jj} and IceCube~\cite{Aartsen:2013jza} experiments show us that the deficit in the muon events  can be understood as the result of oscillation $\nu_{\mu}\to \nu_{\tau}$. Other experiments also show a strong signal for $\nu_\mu$ disappearance. MINOS, for instance, fixed very precisely the scale of oscillations at the value $|\Delta m_{32}^2|~=~\left( 2.41^{+0.09}_{- 0.10} \right)\times~10^{-3}$~eV$^2$ and $\sin^2 2\theta_{23}=0.950^{+0.035}_{-0.036}$~\cite{Adamson:2013whj}. The reactor experiments~\cite{Abe:2012tg,An:2013zwz,Ahn:2012nd} show evidence for electron neutrino disappearance. For instance, the Daya Bay experiment show a signal for oscillations with a scale of $\Delta m_{ee}^2=\left(2.59^{+0.19}_{-0.20}\right)\times 10^{-3}$~eV$^2$  (the $\Delta m_{ee}^2$ parameter is the properly averaged quantity between $\Delta m_{32}^2$ and $\Delta m_{31}^2$~\cite{Minakata:2006gq}) and with amplitude of  $\sin^2 2\theta_{13}= 0.090^{+0.008}_{-0.009}$~\cite{An:2013zwz}.  From solar neutrino experiments~\cite{GonzalezGarcia:2002dz} and the reactor experiment KamLAND~\cite{Gando:2013nba} there is evidence for (anti-)electron neutrino disappearance.  These two signals of oscillations can be explained by $\Delta m^2_{21}=  (7.53\pm0.18) \times 10^{-5}$ eV$^2$ and the large mixing angle  $\tan^2 \theta_{12}=0.436^{+0.029}_{-0.025}$~\cite{Gando:2013nba}, which are associated to what is called the Large Mixing Angle (LMA) solution for the solar neutrino anomaly.

Now that it is established that neutrinos are massive, this also implies that they could decay. The idea of neutrino decay is as old as the idea of  neutrino masses and mixing. The decays of the neutrino in the Standard Model $\nu^{\prime}\to 3\nu$ and  $\nu^{\prime} \to \nu\gamma$ are already too constrained and will not be discussed here (see Ref.~\cite{Mirizzi:2007jd}). An interesting possibility is the scenario where the neutrino decays into another neutrino and a scalar (or Majoron):  $\nu^{\prime} \to \nu+\phi$ decays, where $\phi$ can be a scalar or pseudoscalar massless boson~\cite{Zatsepin:1978iy,Lindner:2001fx}.  These non-radiative decays are from two types: (I) {\it invisible decays}, where neutrinos decay into non-observable final states~\cite{Bandyopadhyay:2001ct,Kelso:2010yz,Lunardini:2010ab,Ando:2003ie,Fogli:2004gy,Barger:1998xk,GonzalezGarcia:2008ru,PalomaresRuiz:2005vf,Adamson:2010wi}; and (II) {\it visible decays}, where the final products contain active neutrinos~\cite{Raghavan:1987uh,Berezhiani:1991vk,Das:2010sd,Eguchi:2003gg,Aharmim:2004uf,Berezhiani:1989za,Lindner:2001th,Ando:2004qe}.

We can parametrize the decay by the ratio of the lifetime parameter $\tau_i$ and the mass $m_i$ for each of the mass eigenstates  $i = 1, 2, 3$. The role of invisible neutrino decay was investigated for the solar neutrino anomaly~\cite{Bandyopadhyay:2001ct,Joshipura:2002fb,Beacom:2002cb} and showed no evidence for the dominance of the decay scenario. From these we can constrain values of the decay parameter, ${\tau_2}/{m_2} > 8.7\times  10^{-5}\, {\rm s/eV}$ at 90\% C.L., where $\tau_2$ and $m_2$ are respectively the lifetime and the highest mass eigenstate in a two generation scenario~\cite{Bandyopadhyay:2001ct}.

In the visible decay scenario, we can search for $\nu_e \to \overline{\nu}_e$ conversion using a pure $\nu_e$ source such as the Sun~\cite{ Raghavan:1987uh}. The null results from the solar $\overline{\nu}_e$ appearance impose a  constrain on the decay parameter:  $\tau_2/m_2 > 6.7 \times 10^{-2}$~s/eV from the KamLAND experiment~\cite{Eguchi:2003gg}. 

Neutrinos produced in supernovas are interesting to investigate for the presence of decays due to the large distance traveled by the neutrino.  From the observation of electron neutrinos from SN1987A~\cite{Hirata:1988ad, Bionta:1987qt}  we should have a lower limit in neutrino lifetime. Otherwise we could not see any signal.  For larger values of the mixing angle, such as the current LMA solution for the solar neutrino anomaly, no constraint is possible~\cite{Frieman:1987as}.  Other possibilities for neutrinos coming from a supernova include the neutrino decay catalyzed by a very dense media~\cite{Berezhiani:1989za,Berezhiani:1991vk} where the matter effects can increase the decay rate of neutrinos. The diffuse supernova neutrinos (neutrinos coming from all past supernova explosions)~\cite{Lunardini:2010ab},
can provide very robust sensitivity in the range of  $\tau/m<10^{10}$~s/eV~\cite{Ando:2003ie, Fogli:2004gy}.

Astrophysical neutrino sources megaparsec away can generate all neutrino flavors. Due to the long distance from the sources we are in the limit $L\to \infty$, where all dependence on the lifetime parameter $\tau_i$ fades away.  However, if we have a precise determination of the ratio of flavors of these neutrinos we can discriminate the case with and without decay~\cite{Meloni:2006gv,Maltoni:2008jr, Baerwald:2012kc, Dorame:2013lka}.

Concerning the accelerator and atmospheric neutrinos we can test the decay scenario of the third generation of neutrino mass eigenstates investigating how the $\tau_3/m_3$ decay parameter changes the $\nu_{\mu} \to \nu_{\mu}$ survival probability~\cite{Barger:1998xk,GonzalezGarcia:2008ru}.
The MINOS experiment made a search for the  decaying  neutrino and constrained the lifetime to $\tau_3/m_3 > 2.1\times 10^{-12}$~s/eV at 90\% C.L., using both neutral and charged current events~\cite{Adamson:2010wi}.
The combined analysis of Super-Kamiokande atmospheric neutrinos with K2K and MINOS accelerator neutrinos show a 90\% C.L. lower bound value of $\tau_3/m_3 > 2.9 \times 10^{-10}$~s/eV~\cite{GonzalezGarcia:2008ru}. 

This article is organized as follows, in Section~\ref{decaymodel} we discuss our neutrino decay scenario. Next, we introduce the $\chi^2$ analysis developed for the neutrino charged and neutral current data of the MINOS experiment in Section~\ref{analysis}. We then present our bounds for the neutrino lifetime based on a charged current analysis (Section~\ref{results_1}) and on a combined charged and neutral current analysis (Section~\ref{results_2}) using MINOS data. In Section~\ref{results_3} we discuss the sensitivity of the T2K experiment for the neutrino decay scenario and present the constraints on neutrino lifetime for analyses using T2K data only and the combined MINOS and T2K data. Section~\ref{sec_maj_models} presents a discussion on the relation of this scenario under Majoron models.

\section{Decay model for neutrinos}
\label{decaymodel}
We are going to introduce the neutrino evolution equation in which the neutrino can decay. This is made by putting an imaginary part  related to the neutrino lifetime, which is the ratio $\alpha_3\equiv m_3/\tau_3$, in the evolution equation. We are going to assume the decay of the heaviest state, $\nu_3\to \nu_s +\phi$, where both final products are invisible. The two-generation system is considered, which is adequate to describe the muon neutrino and anti-neutrino data from MINOS.
The evolution equation is 
\begin{eqnarray}
i \dfrac{d}{dx} \tilde{\nu}
=U\left[\dfrac{\Delta m_{32}^2}{2E} \left(\begin{array}{cc} 
0 & 0\\ 
0 & 1\\ 
\end{array}\right)-i\dfrac{\alpha_3}{2E}
\left(
\begin{array}{cc} 
0 & 0\\ 
0 & 1\\ 
\end{array}
\right)
\right]U^{\dagger} \tilde{\nu}
\label{eq:evolucao}
\end{eqnarray}
where the state  $\tilde{\nu}\equiv \left( \begin{array}{c} 
  \nu_{\mu} \\
\nu_{\tau}\\
\end{array}\right)$, and {\it E} is the neutrino energy. $U$ is the usual rotation matrix,
\begin{eqnarray}
U=\left( \begin{array}{cc} 
  c_{23} & s_{23}  \\
- s_{23}   &  c_{23}  \\
\end{array}\right),
\end{eqnarray}
where $c_{23}\equiv \cos \theta_{23}$ and $s_{23}\equiv \sin \theta_{23}$. The same evolution equation applies for anti-neutrinos as well.

From Eq.~(\ref{eq:evolucao}) we obtain the muon neutrino survival probability as
\begin{align} 
P(\nu_{\mu} \to \nu_{\mu} )  &= \left[\cos^2 \theta_{23}+\sin^2 \theta_{23}e^{-\frac{\alpha_3 L}{2E}}\right]^2  &
\label{eq:decay} \\
& -  4\cos^2 \theta_{23} \sin^2 \theta_{23}e^{-\frac{\alpha_3 L}{2E}} \sin^2\left( \frac{\Delta m_{32}^2 L}{4E} \right)   & \nonumber
\end{align}
where $L$ is the distance traveled by the neutrinos. We can notice that a non-zero decaying parameter $\alpha_3$ changes only the amplitudes: the {\it constant} amplitude (first term of the equation above), and the oscillation amplitude (second term).  Both amplitudes are damped but the oscillation phase does not change.  In the two-$\nu$ standard oscillation probability formula we have the symmetry $\cos^2\theta_{23} \leftrightarrow \sin^2 \theta_{23}$, but in   Eq.~(\ref{eq:decay}) the symmetry is broken, and then we should scan the parameter space of the variable $\sin^2\theta_{23}$ in the range $(0, 1)$. This broken symmetry will appear in our plots later.

The limiting case where the oscillations are induced only by decay, $\Delta m_{32}^2\to 0$, can also be tested. In this case the probability assumes the simple form
\begin{align}
P(\nu_{\mu} \to \nu_{\mu} ) &= \left[\cos^2 \theta_{23}+\sin^2 \theta_{23}e^{-\frac{\alpha_3 L}{2E}}\right]^2 &
\label{eq:onlydecay}
\end{align}
where now we have two free parameters,  the mixing amplitude $\sin^2 \theta_{23}$ and the decay parameter $\alpha_{3}$. Even at this limit we also have an asymmetry between $\sin^2 \theta_{23}$  and  $\cos^2\theta_{23}$.

In the standard neutrino oscillation scenario, the sum of the probabilities over the active states is equal to unity. Then the spectrum of neutral current (NC) events is not effected by active oscillation, which means that the expected number of NC events is the same with or without oscillations. But in the extended scenario involving sterile neutrinos the sum of probabilities, $\sum_\beta P(\nu_\mu~\to~\nu_\beta)$, where $\beta = \mu, \tau$, obviously does not sum up to 1. This is discussed in the general context of non-unitary neutrino evolution in Ref.~\cite{Berryman:2014yoa}.

We can compute the conversion probability for the two-$\nu$ oscillations with decay scenario,
\begin{align} 
P(\nu_{\mu} \to \nu_{\tau} )  &= \cos^2 \theta_{23} \sin^2 \theta_{23} \left[1-e^{-\frac{\alpha_3 L}{2E}}\right]^2  &
\label{eq:decaytau} \\
& +  4\cos^2 \theta_{23} \sin^2 \theta_{23}e^{-\frac{\alpha_3 L}{2E}} \sin^2\left( \frac{\Delta m_{32}^2 L}{4E} \right),   & \nonumber
\end{align}
which implies that
\begin{eqnarray} 
\sum_{\beta=\mu, \tau} P(\nu_{\mu}~\to~\nu_{\beta}) = \cos^2 \theta_{23} + 
\sin^2 \theta_{23} e^{-\frac{\alpha_3 L}{E}}
\label{eq:decaysum} 
\end{eqnarray}
Thus, from the Eq.~(\ref{eq:decaysum}) we observe that there will be an effect on the neutral current interaction events under the oscillations with decay model.

\section{Analysis of MINOS Charged and Neutral Current Data}
\label{analysis}

We have performed a combined analysis using the published data of charged and neutral current MINOS analyses. MINOS is a long-baseline neutrino experiment~\cite{Michael:2008bc} using two detectors and exposed to a neutrino beam produced at Fermilab. The NuMI beam line is a two-horn-focused neutrino beam that can be configured to produce muon neutrinos or anti-neutrinos. The Near Detector is located at Fermilab, around 1 km from the NuMI target and the Far Detector is 735 km far from the target.

The first data set used in our analysis comprises the charged current (CC) contained-vertex neutrino disappearance data~\cite{Adamson:2013whj} using the $\nu_\mu$ enhanced beam with exposure of $10.71 \times 10^{20}$ protons on target (POT), with 23 points of non-equally divided bins of energy up to 14~GeV;  the second is the CC contained-vertex anti-neutrino disappearance data~\cite{Adamson:2013whj} from the $\bar{\nu}_\mu$ enhanced beam with $3.36 \times 10^{20}$ POT, using 12 points for energies up to 14~GeV as well; and the third is the neutral current data~\cite{Adamson:2011ku}, based on $7.07 \times  10^{20}$ POT. The spectrum of NC events is described as a function of a reconstruted energy, $E_{\rm reco}$.  We then use the information from a previous MINOS analysis~\cite{Adamson:2010wi} to separate the NC data into two bins: (a) events with $E_{\rm reco}<3.0$~GeV and (b) events with $3.0<E_{\rm reco}<20.0$~GeV, which have median neutrino energies of 3.1 and 7.9~GeV, respectively.

For the $\chi^2$ calculation we use the following function 
\begin{eqnarray}
\chi^2=\sum_i \dfrac{\left(N_i^{{\rm th}}-N_i^{{\rm data}}\right)^2}{\sigma_i^2}
\label{eq:chi2_calc}
\end{eqnarray}
in bins of energy, where $N_i^{{\rm th}}=N_i^{{\rm mod}}+\beta N_i^{{\rm bg}}$ is the prediction for the theoretical model ($N_i^{{\rm mod}}$) that we are using (e.g. oscillation plus decay scenario), based on the no-oscillation events ($N_i^{{\rm no-osc}}$), and including the background contribution ($N_i^{{\rm bg}}$) adjustable by a parameter $\beta$. $N_i^{{\rm data}}$ is the data from one of the MINOS analyses and $\sigma_i$ is the total error. The background events of the NC data set come from misidentified CC events. The numbers $N_i^{{\rm data}}$, $N_i^{{\rm bg}}$ and $N_i^{{\rm no-osc}}$ were read off from References~\cite{Adamson:2013whj} and~\cite{Adamson:2011ku}.

The total error used for both neutrino and anti-neutrino CC data sets is given by
\begin{eqnarray}
\sigma_i^2~=~\left(\sigma_i^{{\rm data}}\right)^2 + \left(\sigma_i^{{\rm stat,th}}\right)^2 + \left(\sigma_i^{{\rm syst,th}}\right)^2
\end{eqnarray}
where $\sigma_i^{{\rm data}}$ is the total error of the data, $\sigma_i^{{\rm stat,th}}$ and $\sigma_i^{{\rm syst,th}}$ are the statistical and systematic error of the prediction, respectively. We have 
$\sigma_i^{{\rm data}}=\sqrt{\sigma_i^{-}\sigma_i^{+}}$, where $\sigma_i^{-} (\sigma_i^{+})$ is the lower (upper) error bar from the data;  $\sigma_i^{{\rm stat,th}}\equiv \sqrt{N_i^{{\rm th}}}$ and $\sigma_i^{{\rm syst,th}}\equiv 0.04~N_i^{{\rm th}}$~\cite{Adamson:2007gu}.
For the NC data set we do not have the total error of the data events. We then calculate it summing in quadrature the statistical and systematic errors. The former is the square root of the number of data events, and the latter is an estimate based on the systematic error of the extracted expectation of events.

We scan the parameter region in the variables $\tau_3/m_3$, $\sin^2\theta_{23}$ and $\Delta m_{32}^2$ and for the $\beta$ parameter of our function $\chi^2=\chi^2(\tau_3/m_3, \sin^2\theta_{23}, \Delta m_{32}^2, \beta)$. The $\chi^2$ is then marginalized over the nuisance $\beta$ parameter to obtain the effective $\chi^2_{\rm eff}$ as a function of the other parameters
\begin{eqnarray}
\chi^2_{\rm eff}(\tau_3/m_3,s^2_{23}, \Delta m_{32}^2)}= \left.\chi^2(\tau_3/m_3,s^2_{23}, \Delta m_{32}^2,\beta)\right|_{{\rm min}\, \beta
\nonumber
\end{eqnarray}
To have a guess about the range of the decay parameter we can probe, we will assume the argument of the exponential term in Eq.~(\ref{eq:decay}) is of the order of unity, where we can get the most sensitivity. Using the MINOS distance and the energy range of $E_{\nu}=(0.5,~10)$~GeV, we get $\tau_3/m_3 \sim 10^{-13} - 10^{-11}$~s/eV.

\begin{table*}[!t]
\centering
\caption{Comparison between our analyses (MINOS CC only, MINOS CC plus NC, T2K CC only, and MINOS plus T2K) and published analyses by MINOS~\cite{Adamson:2013whj, Adamson:2010wi} and T2K (normal hierarchy)~\cite{Abe:2014ugx}, for no oscillation, standard oscillation, pure decay, and oscillation with decay hypotheses. The values for the square mass difference $|\Delta m^2_{32}|$ and the decay parameter $\tau_3/m_3$ are in $10^{-3}$~eV$^2$ and s/eV, respectively.}

\begin{tabular}{llccccccc}
\hline\hline
{\rm Model}  & & $\chi^2$/d.o.f. & $|\Delta m^2_{32}|$  & $\sin^2 2\theta_{23}$ & ~$\sin^2 \theta_{23}$ &  $\tau_3/m_3$ &  $\tau_3/m_3^{90\% {\rm C.L.}}$\\
\hline
{\rm No oscillation}    & our MINOS CC   & 324.70/35\\ 
				& our T2K CC 	      & 185.92/27\\
\hline                                         
{\rm Standard oscillation} & MINOS Ref.~\cite{Adamson:2013whj}    & NA          & 2.41 & 0.95 &  & \\
                                         & our MINOS CC                                        & 19.80/31 & 2.38 & 0.92 &  & \\ 
                                         & T2K Ref.~\cite{Abe:2014ugx}  	        & NA 	  & 2.51 & 1.00 & 0.51 & \\
                                         & our T2K CC						& 12.66/25 & 2.44 & 0.97 & 0.58 & \\
\hline
{\rm Pure decay} & MINOS Ref.~\cite{Adamson:2010wi}  & 76.4/40   & & 0.96 & 0.60 & $7.3 \times 10^{-13}$ &  \\ 
                           & our MINOS CC                                     & 51.18/31 & &  0.00 & 1.00 & $1.4 \times 10^{-12}$ &  \\
                           & our MINOS CC+NC                              & 69.73/32 & & 0.38 & 0.89 & $1.3 \times 10^{-12}$ &  \\
                           & our T2K CC 				                & 23.11/25 & & 0.86 & 0.69  & $3.5 \times 10^{-13}$ &  \\
                           & our MINOS+T2K 				       & 117.28/59 & & 0.08 & 0.98 & $1.3 \times 10^{-12}$ \\
\hline
{\rm Osc. with decay} & MINOS Ref.~\cite{Adamson:2010wi} & 47.5/39  & NA    & 1.00 & 0.50 & $\infty$  & $> 2.1 \times 10^{-12}$ \\ 
                                   & our MINOS CC        & 19.28/30 & 2.30 & 0.93 & 0.63 & $1.2  \times 10^{-11}$ & $>2.0  \times 10^{-12}$ \\
                                   & our MINOS CC+NC  & 19.88/31 & 2.34 & 0.96 & 0.60 & $\infty$ & $>2.8  \times 10^{-12}$ \\
                                   & our T2K CC              & 7.33/24  & 2.46 & 1.00 & 0.49 & $ 1.6\times 10^{-12}$ & $ >7.8\times 10^{-13}$ \\
                                   &                                    & &  &  &  & &  {\rm and}  $<8.3\times 10^{-12}$ \\
                                   & our MINOS+T2K                        & 32.48/58 & 2.34 & 0.97 & 0.42 & $8.5  \times 10^{-12}$   & $>2.8  \times 10^{-12}$ \\
\hline \hline
\end{tabular}
\label{tab:parameters}
\end{table*}

\section{Results}
\subsection{MINOS Charged Current Analysis}
\label{results_1}

To test the correctness of our $\chi^2$ analysis, we first consider the standard oscillation scenario, corresponding to the limit $\alpha_3 \to 0$ in Eq.~(\ref{eq:decay}).  The results shown here are from neutrino and anti-neutrino charged current data obtained from Ref.~\cite{Adamson:2013whj}. However, we also used our calculation on the neutrino charged current data from Ref.~\cite{Adamson:2011ig}. For both data sets we obtain good concordance with the MINOS allowed region and best fit.

In Table~\ref{tab:parameters} we show the consistency between the best fit parameters obtained by our standard oscillation analysis (accelerator data only) and MINOS~\cite{Adamson:2013whj} (which includes accelerator and atmospheric data). We obtain that the best fit of the oscillation parameters is for non-maximal mixing angle, $\sin^2 2\theta_{23} =  0.92$ and for $|\Delta m^2_{32}|=2.38\times 10^{-3}$~eV$^2$. The $\chi^2$ of the best fit point is 19.80 for 31 degrees of freedom, which corresponds to a G.O.F. (Goodness of Fit) of 94.0\%.
The $\Delta \chi^2$ for the no-oscillation hypothesis compared to the standard oscillation scenario,
\begin{eqnarray}
\Delta \chi^2\equiv \chi^2_{\rm no-oscillation}-\chi^2_{\rm standard\_oscillation}
\nonumber
\end{eqnarray}
is equal to 304.9, which means that we can exclude the no-oscillation hypothesis with remarkable precision.

\begin{figure}[tttt!]
\centering
\includegraphics[scale=0.4]{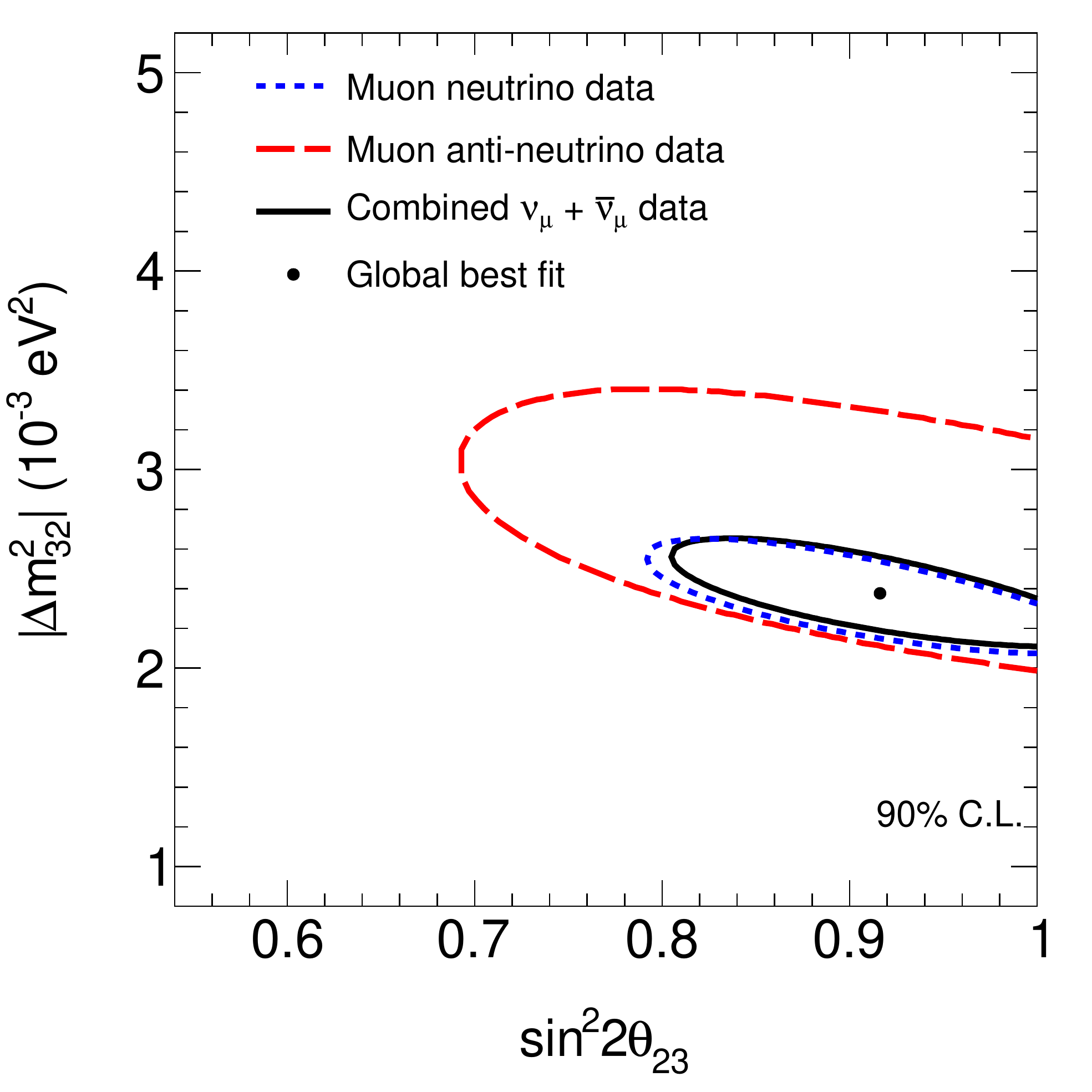}
\caption{Allowed regions at the 90\% C.L. in the $\sin^2 2\theta_{23}$ -- $\Delta m^2_{32}$ plane from standard oscillation fits of the neutrino and anti-neutrino data separately (dashed curves) and of the combined analysis (solid curve).}
\label{fig:AR_2D_dm2_s2t_90CL}
\end{figure}

The allowed regions in the ($\sin^2 2\theta_{23}$ -- $\Delta m^2_{32}$) plane for neutrinos and anti-neutrinos under the standard oscillations model is shown in Fig.~\ref{fig:AR_2D_dm2_s2t_90CL}. The anti-neutrino data is compatible with the neutrino data but does not contribute significantly to improve the region of parameters in the combined CC analysis due to its small statistics compared to the neutrino one. Our results show that we are able to reproduce the allowed regions of square mass difference and mixing angle of the MINOS analyses.

\begin{figure}[tttt!]
\centering
	\includegraphics[scale=0.4]{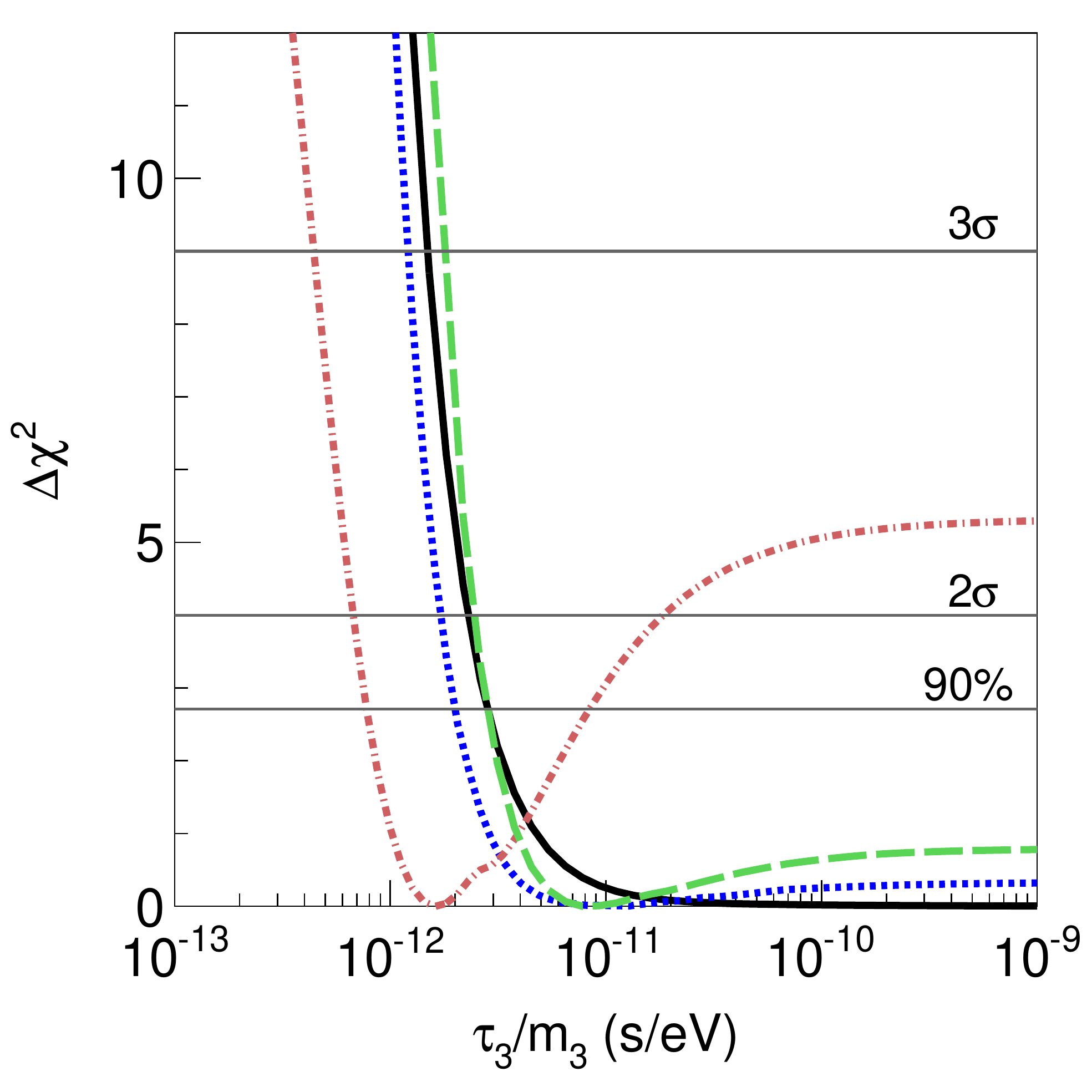}
\caption{Projections of $\Delta \chi^2$ as a function of $\tau_3 / m_3$ for the oscillation with decay model using the MINOS CC analysis (dotted curve), the MINOS combined CC with NC analysis (solid curve), the T2K CC analysis (dash-dotted curve), and the combined MINOS and T2K analysis (dashed curve). The ranges of allowed parameter lie below the horizontal lines at 90\%, 2$\sigma$ and 3$\sigma$ confidence levels.
}
\label{fig:tau}
\end{figure}

Having confirmed the correctness of the analysis, we use it to test the oscillation with decay scenario. The procedure is the same as in the standard oscillation case, but now we have three free parameters,  $\Delta m^2_{32}$, $\sin^2 \theta_{23}$ and $\tau_3/m_3$.  For the combined (neutrino and anti-neutrino) CC analysis under the oscillation with decay framework we find
a finite  value for the best fit of $\tau_3/m_3$~(see Table~\ref{tab:parameters}),
\begin{eqnarray}
  \tau_3/m_3 = 1.2 \times 10^{-11} ~{\rm s/eV}.
\label{b.f.II}
\end{eqnarray}
The effect of a finite value of  $\tau_3/m_3$ results in a slightly lower value of $\Delta m^2_{32}$ than the one obtained for standard oscillation.
The $\chi^2$ obtained is 19.28 for 30 degrees of freedom, which corresponds to a  G.O.F. of 93.4\%. The pure decay hypothesis is also considered in our CC analysis resulting in a $\chi^2$ of 51.18 for 31 degrees of freedom, excluding this hypothesis at the 5.6 standard deviation level.

\begin{figure}[tt!]
\centering
\includegraphics[scale=.4]{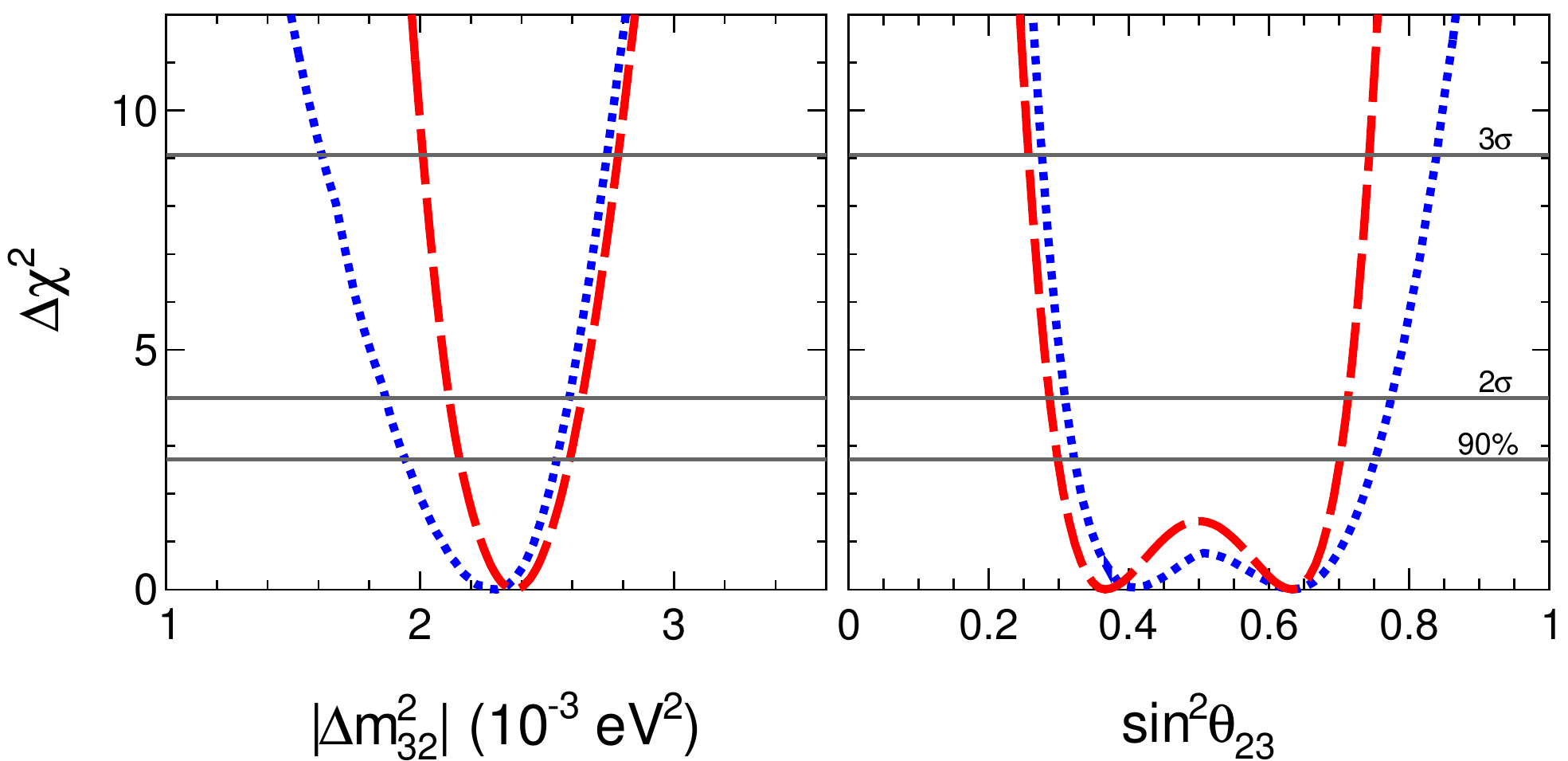}
\caption{Projections of $\Delta \chi^2$ as a function of $|\Delta m_{32}^2|$ (left) and $\sin^2 \theta_{23}$ (right) for the standard oscillation (dashed curves) and oscillation with decay model for the CC only (dotted curves). The ranges of allowed parameter lie below the horizontal lines at 90\%, 2$\sigma$ and 3$\sigma$ confidence levels.
}
\label{fig3:tot}
\end{figure}

The projections of $\Delta \chi^2$ for each parameter are obtained minimizing the function $\chi^2 = \chi^2(\tau_3/m_3,~\sin^2\theta_{23},~\Delta m_{32}^2)$ accordingly to the parameters. We then use the function $\Delta \chi^2\equiv  \chi^2-\chi^2_{{\rm min}}$, where $\chi^2_{{\rm min}}$ is the global minimum value for the oscillation with decay scenario. Next, we present the projections of $\Delta \chi^2$ as a function of $\tau_3/m_3$, $\Delta m^2_{32}$ and $\sin^2\theta_{23}$. 

The one-dimensional projection for the $\tau_3 / m_3$ parameter, shown in Fig.~\ref{fig:tau} (dotted curve), allows us to get a lower bound on $\tau_3 / m_3$ or equivalently an upper bound on $\alpha_3$. The allowed values for the neutrino decay lifetime is
\begin{eqnarray}
\tau_3/m_3 > 2.0 \times 10^{-12}~{\rm s/eV}
\end{eqnarray}
at the 90\% C.L. (Table~\ref{tab:parameters}).  We can compare the value we obtain for the oscillation with decay model to the one from MINOS~\cite{Adamson:2010wi}, which used both charged and neutral current data in their analysis, but with lower statistics than the extracted data used here. We see that our combined CC result does not improve the MINOS lower limit which is $\tau_3/m_3 > 2.1 \times 10^{-12}$ s/eV.

An interesting behavior appears in Fig.~\ref{fig3:tot} (left), where we compare the allowed values for $\Delta m^2_{32}$  with and without decay. When we include a finite value for the $\tau_3/m_3$ parameter, smaller values of $\Delta m^2_{32}$  are allowed for a certain confidence level. Then, under this scenario, we can explain the muon disappearance signal seen by MINOS as due partially to oscillation and partially to decay, since the limit $\Delta m^2_{32} \to 0$ in Eq.~(\ref{eq:decay}) corresponds to the pure decay model.

\begin{figure}[tt]
\centering
\includegraphics[scale=.4]{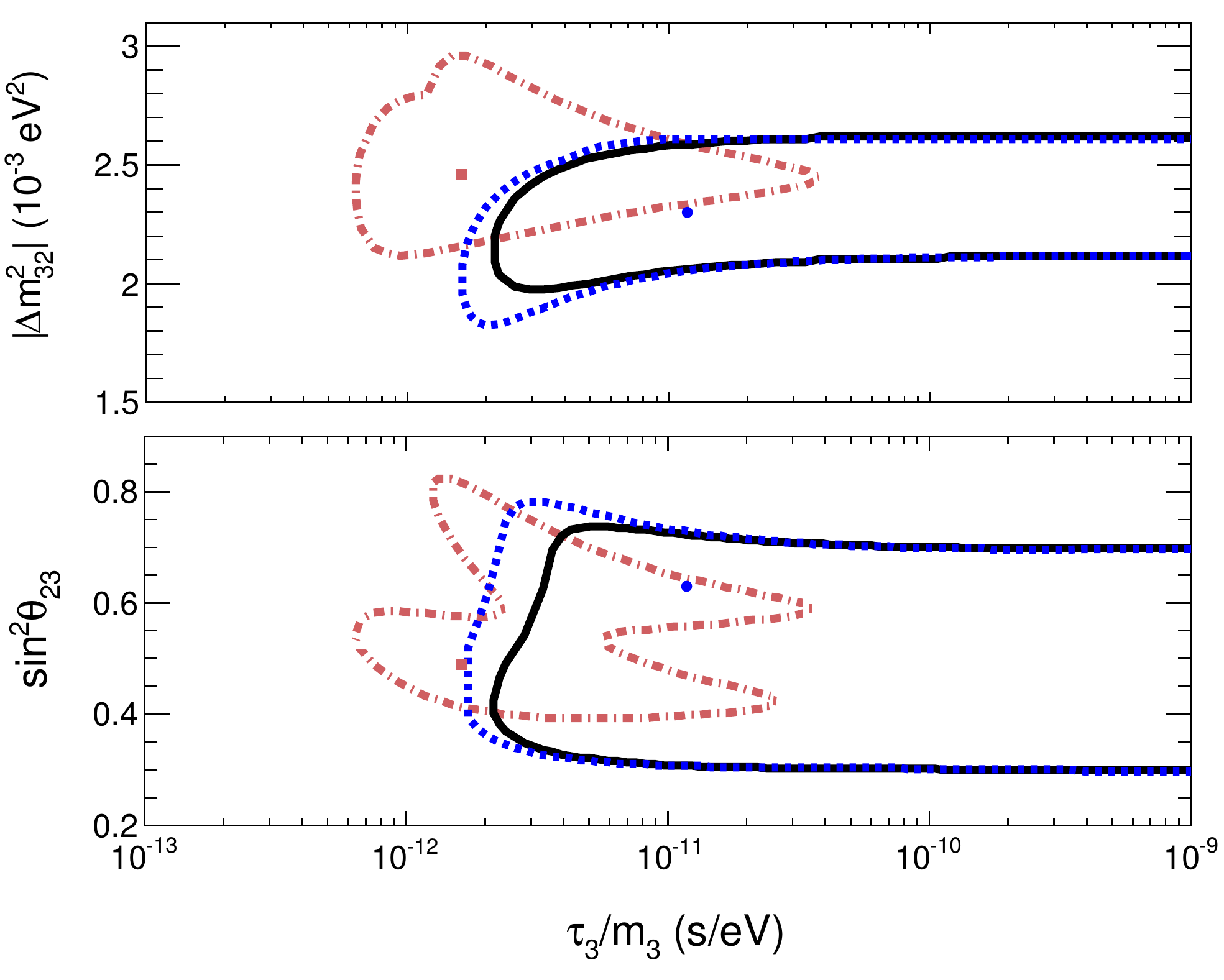}
\caption{Allowed regions at 90\% C.L. of $\tau_3/m_3$ versus $|\Delta m^2_{32}|$ (top) and $\tau_3/m_3$ versus $\sin^2 \theta_{23}$ (bottom) for the oscillation with decay scenario using the MINOS CC data only (dotted curves), the MINOS combined CC with NC data (solid curves), and T2K CC data (dash-dotted curves). The best fit points are for the MINOS CC data (circles) and for the T2K CC data (squares).}
\label{fig4:tau_dm2_st2}
\end{figure}

The broken symmetry, $\cos^2\theta_{23} \leftrightarrow \sin^2 \theta_{23}$ mentioned after Eq.~(\ref{eq:decay}), is manifested in Fig.~\ref{fig3:tot} (right), where the curve is broader for larger values of $\sin^2 \theta_{23}$. This effect can be understood investigating the coefficient on the first term of Eq.~(\ref{eq:decay}). For $\sin^2 \theta_{23}>\cos^2 \theta_{23}$ the decay term is more relevant, and for the opposite case the decay contribution is suppressed.

We also present two-dimensional projections of the allowed three-dimensional region after normalization with respect to the undisplayed parameter. Figure~\ref{fig4:tau_dm2_st2} shows the 90\% C.L. allowed region in the ($\tau_3/m_3$ -- $\Delta m^2_{32}$) and ($\tau_3/m_3$ -- $\sin^2 \theta_{23}$) planes for the oscillation with decay scenario (dotted curves) and our best fit point (circles) for the MINOS CC data only. We can observe that for smaller values of $\tau_3/m_3$, when the effects of the decay are larger, the contours allow smaller values of $\Delta m^2_{32}$ and higher values of $\sin^2 \theta_{23}$ (the same behaviour shown in Fig.~\ref{fig3:tot}).

\begin{figure}[tt]
\centering
	\includegraphics[scale=0.4]{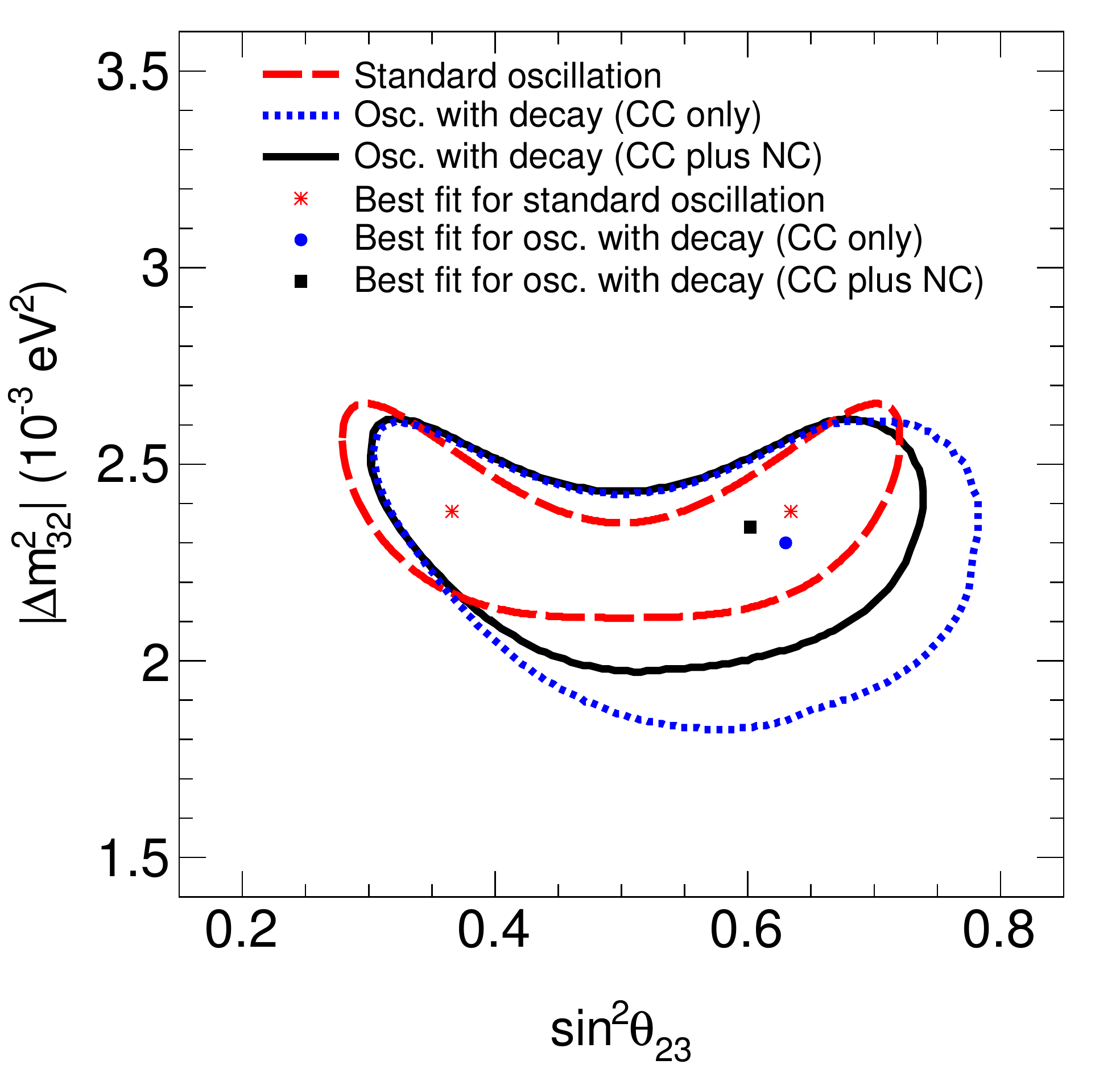}
\caption{Best fit points and allowed regions at 90\% C.L. of $\sin^2 \theta_{23}$ versus $|\Delta m^2_{32}|$ for the standard oscillation (dashed curve) and for the oscillation with decay scenario using the MINOS CC data only (dotted curve) and the MINOS combined CC with NC data (solid curve).}
\label{fig5:deltasine}
\end{figure}

The 90\% C.L. allowed regions for the oscillation parameters,  $\Delta m^2_{32}$ and $\sin^2 \theta_{23}$, is presented in Fig.~\ref{fig5:deltasine} both  with and without decay. In agreement with the information shown in the discussion of Figs.~\ref{fig3:tot} and \ref{fig4:tau_dm2_st2} we can see that the decay allows a region of smaller values of  $\Delta m^2_{32}$ and larger values of $\sin^2 \theta_{23}$ than the standard oscillation model.

The best fit points are also shown in Fig.~\ref{fig5:deltasine} for both models (with and without decay). The standard oscillation analysis obviously shows two possible values of $\theta_{23}$ due to the symmetry $\cos^2\theta_{23} \leftrightarrow \sin^2 \theta_{23}$. Since the oscillation with decay model does not manifest such a symmetry we find that the best fit point for this scenario is in the $\theta_{23} > 45^\circ$ octant, with a value of $\sin^2 \theta_{23} = 0.63$.

\begin{figure*}[tt!]
\centering
\includegraphics[scale=0.40]{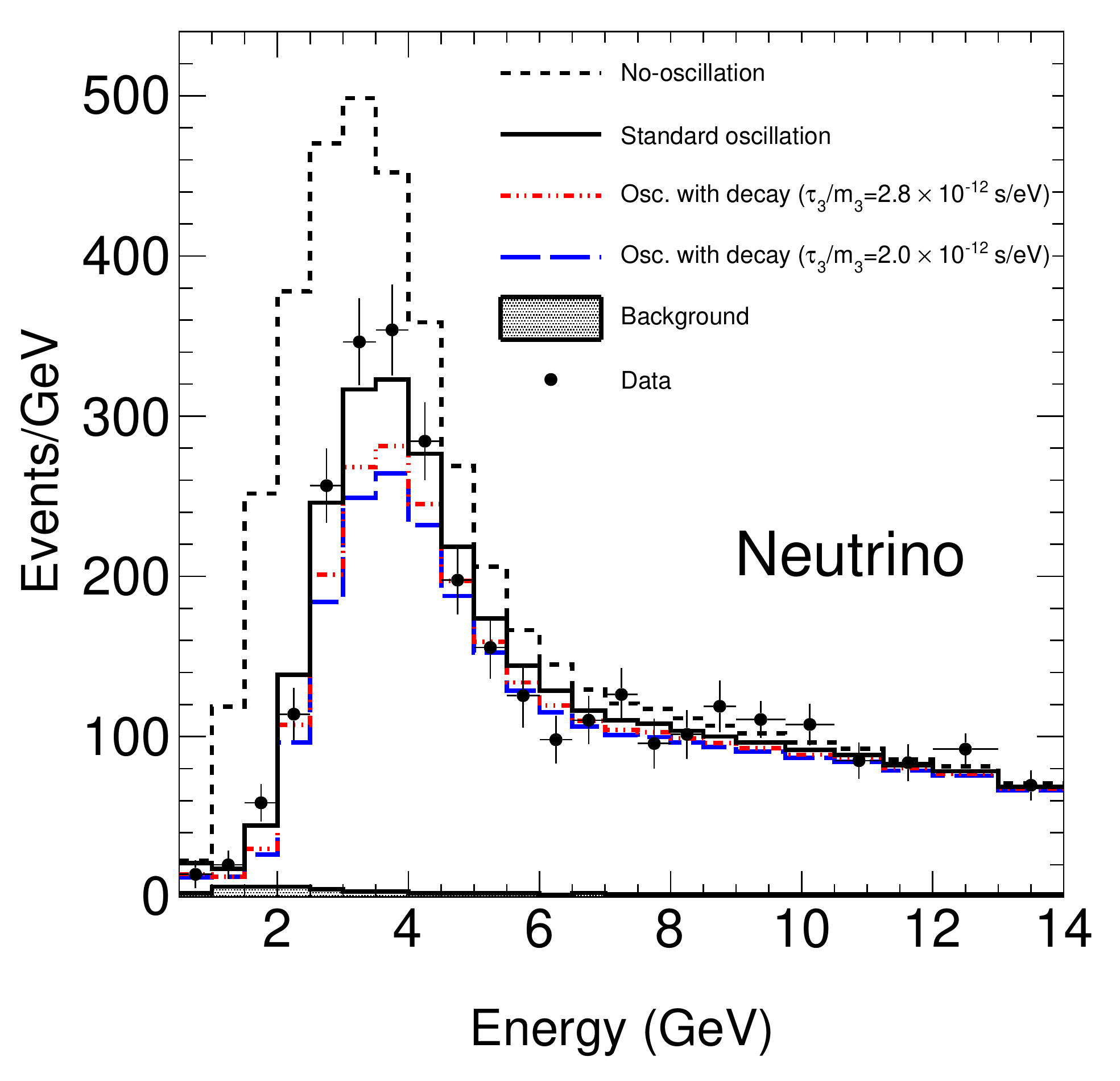} ~~%\label{fig:hist_numu}
\includegraphics[scale=0.40]{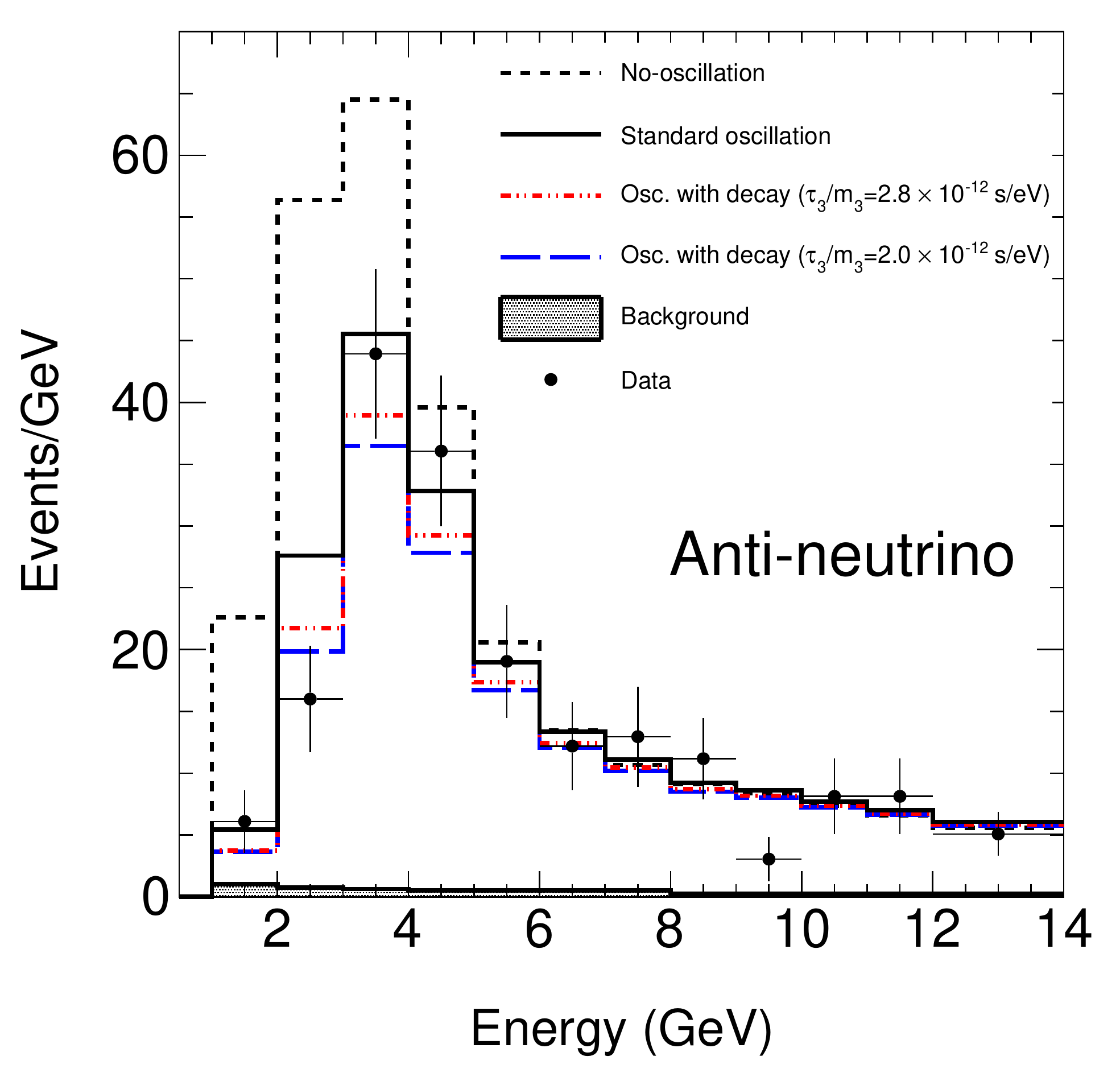} %\label{fig:hist_numubar}
\caption{The extracted charged current neutrino (left) and anti-neutrino (right) spectra of MINOS shown together with the curves for the following hypotheses: (i) no-oscillation, (ii) standard oscillation at the best fit parameters, and (iii) oscillation with decay at the best fit for $\Delta m^2_{32}$ and $\sin^2 \theta_{23}$ and the 90\% C.L. value of $\tau_3/m_3$ for the MINOS CC data only ($\tau_3/m_3 = 2.0 \times 10^{-12}$~s/eV) and for the MINOS combined CC with NC data ($\tau_3/m_3 = 2.8\times 10^{-12}$~s/eV). The hatched areas show the background from neutral current events.}
\label{fig:hist_numu}
\end{figure*}

\subsection{MINOS Combined Charged and Neutral Current Analysis}
\label{results_2}

In this section we present the results obtained for the MINOS combined charged and neutral current data. We expect an effect on the neutrino lifetime due to the inclusion of the neutral current data in the analysis as shown in Eq.~\ref{eq:decaysum}.
First, we calculate the ratio
\begin{eqnarray}
R_i = \frac{N_i^{\rm data} - N_i^{\rm bg}}{N_i^{\rm th}}
\end{eqnarray}
for each bin of reconstructed energy extracted from the NC data, where $N^{\rm data}$, $N^{\rm bg}$, and $N^{\rm th}$ are the number of data, background and expected events, respectively. The ratios we obtain are in good agreement with the ones from MINOS~\cite{Adamson:2011ku}.

Under the oscillation with decay model the best fit point for the combined CC and NC analysis can be found in Table~\ref{tab:parameters}. The $|\Delta m^2_{32}|$ and $\sin^2 \theta_{23}$ values are consistent with the ones for standard oscillations. We obtain a neutrino lifetime $\tau_3/m_3 \to \infty$, which corresponds to the case of no decay. If we compare this result to the best fit of the CC analysis, the decay effect becomes less relevant due to the inclusion of NC data, implying that the best fit value of $|\Delta m^2_{32}|$ increases (from $2.30 \times 10^{-3}$~eV$^2$ to $2.34 \times 10^{-3}$~eV$^2$).

The $\chi^2$ of the best fit for the combined CC and NC analysis is 19.88 for 31 degrees of freedom resulting in a G.O.F. of 93.8\%. The pure decay model is also used in our analysis, being excluded at the 7.1 standard deviation level. Our best fit value of $\tau_3/m_3$ for the pure decay scenario is consistent to the one from MINOS~\cite{Adamson:2010wi}.

The one-dimensional projection for the $\tau_3/m_3$ parameter in the combined CC and NC analysis is shown in Fig.~\ref{fig:tau} (solid curve). We find a lower limit of
\begin{eqnarray}
\tau_3/m_3 > 2.8 \times 10^{-12}~{\rm s/eV}
\label{constrain-cncc}
\end{eqnarray}
at the 90\% C.L., which improves the MINOS~\cite{Adamson:2010wi} limit. This constraint is the best so far for long-baseline experiments.

The two-dimensional projections in Figs.~\ref{fig4:tau_dm2_st2} and \ref{fig5:deltasine} show the effect of including the NC data into our analysis.  From Eq.~(\ref{eq:decaysum}) we know that the effect of decay for the NC sample is stronger for large values of $\sin^2 \theta_{23}$ than for small ones. And since the combined CC and NC analysis does not suggest evidence for decay ($\tau_3/m_3 \to \infty$), this explains the shrink of the large values of $\sin^2 \theta_{23}$ in the allowed region of Fig.~\ref{fig4:tau_dm2_st2} (bottom). The lower values of $\Delta m_{23}^2$ in the allowed region at the top plot of Fig.~\ref{fig4:tau_dm2_st2} was cut due to its correlation with the larger values of $\sin^2 \theta_{23}$.

In Fig.~\ref{fig5:deltasine} we can see that the inclusion of NC data into our analysis decreases the asymmetry between $\cos^2 \theta_{23}$ and $\sin^2 \theta_{23}$ which is present in the CC only analysis.  For the combined CC and NC analysis, larger values of  $\sin^2 \theta_{23}$ are also excluded by the same reason pointed out in the discussion of the Fig.~\ref{fig4:tau_dm2_st2}.

The extracted MINOS spectrum data for CC neutrinos and anti-neutrinos are shown in Fig.~\ref{fig:hist_numu}. In both plots we show the curves for the no-oscillation hypothesis, the best fit of standard oscillations, and the background from NC events. In addition, we also show the curves for the oscillation with decay scenario at the best fit for $\Delta m^2_{32}$ and $\sin^2 \theta_{23}$ and using the 90\% C.L. value of the $\tau_3/m_3$ parameter for the CC data only ($\tau_3/m_3 = 2.0 \times 10^{-12}$~s/eV) and for the combined CC with NC data ($\tau_3/m_3 = 2.8\times 10^{-12}$~s/eV).
The reason for using the 90\% C.L. values of $\tau_3/m_3$ is that the curves for the best fit parameters show no significant difference if compared to the standard oscillations curve. We can observe that the curve using the higher value of $\tau_3/m_3$, which corresponds to smaller effect of decay, is closer to the one for standard oscillations. We also observe that the inclusion of the decay scenario into the oscillation model worsens the fit for both neutrino and anti-neutrino analyses.

\subsection{Combined MINOS and T2K Analysis}
\label{results_3}

We investigate the impact on our analysis of the $\nu_\mu$ charged current data from the T2K experiment~\cite{Abe:2014ugx}. T2K is a long-baseline neutrino experiment which uses two detectors exposed to a neutrino beam produced at J-PARC ring. The Near Detector is located around 280 m downstream of the neutrino production target and the off-axis Far Detector is located 295 km far from the target.  The data we used here has an integrated POT of $6.57 \times 10^{20}$ from the $\nu_\mu$ disappearance analysis.
  
\begin{figure}[tttt!]
\centering
\includegraphics[scale=0.44]{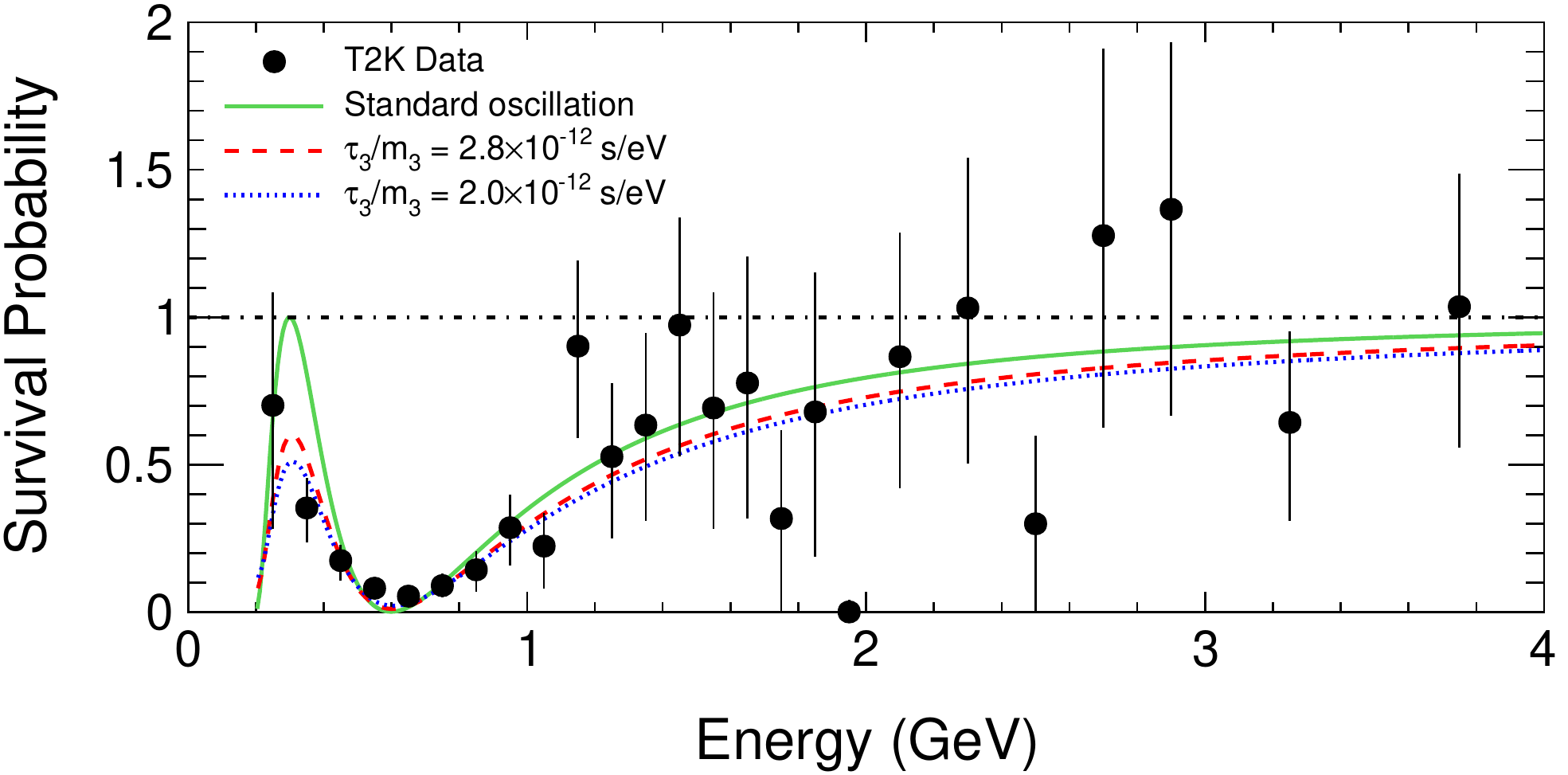}
\caption{The $\nu_\mu$ survival probability for some values of $\tau_3/m_3$, where the solid curve is at the standard oscillation limit, and the dotted/blue (dashed/red) curve is at the constraint from the MINOS charged current analysis (the MINOS combined charged and neutral analysis) under the oscillation with decay scenario. For the  mixing values we use the values from  the  T2K experiment (normal hierarchy)~\cite{Abe:2014ugx}: $\Delta m_{32}^2=2.51\times 10^{-3}$~eV$^2$ and $\sin^2 2 \theta_{23}=1$. The points are the ratio between the experimental T2K data over the prediction with no oscillation. The horizontal line is the prediction for the no oscillation hypothesis. 
}
\label{fig:sensitivity}
\end{figure}

Figure~\ref{fig:sensitivity} shows the $\nu_\mu$ survival probability as a function of the neutrino energy using T2K data for some values of neutrino lifetime under the oscillation with decay scenario. Considering the 90\% C.L. limit for $\tau_3/m_3$ from our MINOS CC and NC analysis (dashed curve) we observe an effect on the probability in the first two T2K data points when comparing to the standard oscillation probability (solid curve). The region at the oscillation minimum, where the errors are smaller than in the other regions, shows very small difference for the scenarios with and without decay. Since T2K data is still dominated by statistical errors their sensitivity for the oscillation with decay scenario would certainly benefit with increasing statistics.
 
Nevertheless, we use the charged current data from the T2K experiment and perform a similar $\chi^2$ analysis as made for the MINOS experiment, shown in Eq.~(\ref{eq:chi2_calc}). The number of data ($N_i^{{\rm data}}$) and expected no-oscillation ($N_i^{{\rm no-osc}}$) events were read off respectively from References~\cite{Abe:2014ugx} and~\cite{Giganti:2014}. Using the T2K data only we have found that the no-oscillation scenario is highly disfavored as shown in Table~\ref{tab:parameters}. Under the standard oscillation scenario, we could reproduce the 90\% C.L. allowed region and obtain best fit values consistent to the T2K ones (Table~\ref{tab:parameters}). The scenario of pure decay for the T2K data is disfavoured compared to standard oscillations, but still being allowed with a G.O.F. of 57.1\%. In the more general scenario of decay and oscilations we have found that T2K data prefers non-zero values of decay parameter, $\tau_3/m_3=1.6\times 10^{-12}$ s/eV with  $|\Delta m_{32}^2|= 2.46 \times 10^{-3}$~eV$^2$ and $\sin^2 2\theta_{23}=1.00$. This data favours the oscillation with decay scenario with a $\chi^2$/d.o.f. = 7.33/24 over the standard oscillation scenario, $\chi^2$/d.o.f. = 12.66/25, mainly due to the energy bin centered at 0.35 GeV (second data point in Fig.~\ref{fig:sensitivity}).

We also test the combined MINOS and T2K data analysis under the pure decay scenario, resulting in its exclusion at the 10.5 standard deviation level. The analysis of the combined data considering the oscillation with decay model results in a finite best fit value for the neutrino lifetime, $\tau_3/m_3 = 8.48 \times 10^{-12}$~s/eV, with the oscillation parameters $|\Delta m^2_{32}|=2.34\times 10^{-3}$~eV$^2$ and $\sin^2 2\theta_{23} =  0.97$ (Table~\ref{tab:parameters}).

Figure~\ref{fig:tau} shows the projections of $\Delta \chi^2$ as a function of $\tau_3 / m_3$ for (I) the T2K only and (II) the combined MINOS and T2K analyses under the oscillation with decay scenario. We already observed in that figure that the solid curve, from the MINOS combined charged and neutral current analysis, shows a non-finite best fit value for $\tau_3/m_3$, as described previously. The dash-dotted curve which is from the T2K data only analysis shows a 90\% C.L. range of the neutrino lifetime, $7.8 \times 10^{-13}<\tau_3/m_3<8.3 \times 10^{-12}$ s/eV. This result is consistent to the closed allowed regions at 90\% C.L. of $\tau_3/m_3$ versus $|\Delta m^2_{32}|$ and $\tau_3/m_3$ versus $\sin^2 \theta_{23}$ for the oscillation with decay scenario (dash-dotted curve) shown in Fig.~\ref{fig4:tau_dm2_st2}.

In Fig.~\ref{fig:tau} we also see the curve (dashed) from the combined MINOS and T2K analysis, that results in a limit of
\begin{eqnarray}
\tau_3/m_3 > 2.8 \times 10^{-12}~{\rm s/eV}
\end{eqnarray}
at the 90\% C.L., which does not improve the contraint obtained by our MINOS analysis. The results due to the inclusion of T2K in our analysis are in agreement with what we expected from the discussion of Fig.~\ref{fig:sensitivity}.

\section{Majoron models for neutrino decay}
\label{sec_maj_models}

The phenomenological model of neutrino decay that we defined by Eq.~(\ref{eq:evolucao}) can be accommodated in well motivated  Majoron models. In general grounds the effective  Lagrangian for neutrino decay can be written as
 \begin{eqnarray}
-{\cal L}=g_{ij}\bar{\nu}_i\nu_j \phi +h_{ij}\bar{\nu}_i\gamma_5 \nu_j \phi+\, {\rm h.c.}
\label{effective}
\end{eqnarray}
where {\it g} and {\it h} are the Majoron-neutrino couplings in mass basis with  scalar (pseudo-scalar)  massless $\phi$ boson. We can compute the neutrino lifetime for the decay $\nu_i \to \nu_j + \phi$, where $\nu_i$ is a neutrino eigenstate.  Assuming that the third mass eigenstate decay into a lightest  state, we have the lifetime in the laboratory system, for $m_3 \gg m_{\rm light}$, given by
\begin{eqnarray}
\Gamma_{\nu_3}^{\rm lab}=\left(\dfrac{g^2+h^2}{32\pi}\right)\dfrac{m_3^2}{E_3}
\label{eq:majoron}
\end{eqnarray}
We can use the Eq.~(\ref{eq:majoron}) to compute the decay parameter ${\tau_3}/{m_3}$ as
\begin{eqnarray} 
\dfrac{\tau_3}{m_3}=\dfrac{1}{E_3\Gamma_{\rm lab}}
\label{gammalab}
\end{eqnarray}
If we use the constraint obtained in our CC (CC+NC) analysis, $\tau_3/m_3 > 2.0 (2.8) \times 10^{-12}~{\rm s/eV}$ at 90\% C.L., into the Eq.~(\ref{gammalab}) we can get an upper limit for  Majoron-neutrino coupling constant of the order of 
\begin{eqnarray}
\dfrac{\sqrt{g^2+h^2}}{10^{-1}}<  \dfrac{1.7 (1.5)~\rm eV}{m_3}
\end{eqnarray}

\def\SM{$\mathrm{SU(3)_c \otimes SU(2)_L \otimes U(1)_Y}$ }
\def\SMextra{$\mathrm{SU(3)_c \otimes SU(2)_L \otimes U(1)_Y \otimes U(1)_H}$ }
\def\SMtres{$\mathrm{SU(3)_c \otimes SU(3)_L \otimes U(1)_N}$ }

The effective Lagrangian for neutrino decay shown in Eq.~(\ref{effective}) can be embedded in different extensions of Standard Model. The general trend is to have the inclusion of new scalar particles in different representations with non-universal couplings between the different families and also the addition of new sterile neutrino states. For instance, Ref.~\cite{Dorame:2013lka} presents a model with an \SMextra  symmetry that included as new fields one extra singlet scalar boson and three right-handed neutrinos.   This model has the $\mathrm{U(1)_H}$ assignments that are family-dependent and therefore the Majoron-neutrino couplings and the mass basis are not proportional to each other, making possible the neutrino decay~\cite{Dorame:2013lka}.

Another example is the model with an \SMtres symmetry that can  have a see-saw scale at TeV energies, when we include two scalars with a singlet and doublet scalar added to this enlarged gauge symmetry~\cite{Cogollo:2008zc}.  Due to anomaly cancellation this model has already family-dependent couplings that can induce neutrino decays. In both examples we can accommodate the Majoron-neutrino couplings to be below the upper limit obtained in this analysis.

\section{Conclusions}

The present scenario of standard neutrino oscillation observed by different experiments shows a strong case for non-zero neutrino masses and mixing. A direct consequence of non-zero neutrino masses is that the neutrino can decay, but usually with much longer lifetimes for decays like $\nu^{\prime}\to 3\nu$ or $\nu^{\prime}\to \nu+\gamma$. The only expectation to observe sizeable effects from neutrino decays is from $\nu^{\prime}\to \nu_s+\phi$, where the decay products are sterile states and the scalar or pseudo-scalar massless boson $\phi$.

We use the more updated charged and neutral current data of the MINOS experiment for accelerator produced neutrinos and anti-neutrinos to study the impact of a non-zero decay parameter, $\alpha_3=m_3/\tau_3$, in the $\nu_{\mu}\to\nu_{\mu}$ channel. Using the charged current data only we have found that the best fit scenario is for a non-zero $\alpha_3$ parameter, corresponding to a neutrino lifetime $\tau_3/m_3 = 1.2 \times 10^{-11}$~s/eV.

We have found that the 90\% C.L. range  for $|\Delta m^2_{32}|$ is $(1.95 - 2.54) \times 10^{-3}$~eV$^2$ and for the mixing angle is $\sin^2 \theta_{23} = (0.32 - 0.75)$ in the oscillation with decay scenario. We could understand the effect of the non-zero decaying parameter for the CC data analysis, implying smaller values of $\Delta m^2_{32}$ and larger values of $\sin^2 \theta_{23}$ than the ones in standard oscillation scenario.

Based on the MINOS combined charged and neutral current analysis the best fit for the oscillation with decay model indicates a zero value for the $\alpha_3$ parameter, that is equivalent to the standard oscillation model. Using the combined analysis we could improve the previous constraint on the allowed neutrino decay lifetime $\tau_3/m_3>2.8 \times 10^{-12}$~s/eV at the 90\% C.L., which is the best limit based only on accelerator produced neutrino data. The inclusion of NC data, which is compatible to standard oscillations, makes the decay scenario more constrained than in the CC analysis only.

We showed that the effect due to the T2K data into our analysis of the MINOS data would be small. In fact, the combined MINOS and T2K analysis did not improve the 90\% C.L. limit on the $\tau_3/m_3$. However, the combined analysis results in a range of finite values for the neutrino lifetime with low significance.

The phenomenological model used in our analysis can be connected with the Majoron models that include also sterile neutrino states.  Our constraint on the neutrino lifetime can be translated  into  the Majoron-neutrino  coupling upper bound as $\dfrac{\sqrt{g^2+h^2}}{10^{-1}}<  \dfrac{1.5~\rm eV}{m_3}$ using the combined charged and neutral current analysis.

\section*{Acknowledgments}
We would like to thank C. Castromonte for helping on computational tasks and M. Goodman for valuable discussion and careful reading of the manuscript. A.~L.~G.~G. was supported by CAPES during his Msc.\ degree project. O.~L.~G.~P. thanks the support of funding grant  2012/16389-1, S\~ao Paulo Research Foundation (FAPESP) and ICTP for hospitality.  And R.~A.~G. thanks the support of Fulbright/CAPES, and the hospitality at Argonne National Laboratory.

%% If you have bibdatabase file and want bibtex to generate the
%% bibitems, please use
%%
%\bibliographystyle{elsarticle-num} 
%\bibliographystyle{atlasBibStyleWoTitle}
\bibliographystyle{apsrev}
%%  \bibliography{<your bibdatabase>}
\bibliography{paper_phys_lett_b}

%% else use the following coding to input the bibitems directly in the
%% TeX file.

\end{document}